\documentclass{ws-ijmpd}
\usepackage[super,compress]{cite}
\usepackage[dvips,breaklinks]{hyperref}
\hypersetup{colorlinks,urlcolor=black,citecolor=black,linkcolor=black,filecolor=black}
\usepackage{breakurl}

\begin{document}
\markboth{Ya-zheng Tao}
{The estimation of far-field wavefront error of tilt-to-length distortion coupling}

%%%%%%%%%%%%%%%%%%%%% Publisher's Area please ignore %%%%%%%%%%%%%%%
%
\catchline{}{}{}{}{}
%
%%%%%%%%%%%%%%%%%%%%%%%%%%%%%%%%%%%%%%%%%%%%%%%%%%%%%%%%%%%%%%%%%%%%

\title{The estimation of far-field wavefront error of tilt-to-length distortion coupling in space-based gravitational wave detection}
\author{$^{1}$Ya-Zheng Tao, $^{2,4}$Hong-Bo Jin\footnote{Corresponding author: hbjin@bao.ac.cn}, $^{1,3,4}$Yue-Liang Wu}
\address{
$^1$University of Chinese Academy of Sciences, No.19(A) Yuquan Road, Shijingshan District, Beijing, P.R.China 100049\\
$^2$National Astronomical Observatories, Chinese Academy of Sciences, 20A Datun Road, Chaoyang District, Beijing, P.R.China 100101\\
$^3$Institute of Theoretical Physics, Chinese Academy of Sciences, No.55 Zhongguancun East Road, Haidian District, Beijing, P.R.China 100190\\
$^4$Hangzhou Institute for Advanced Study, University of Chinese Academy of Sciences, Hangzhou, Zhejiang, P.R.China 310024}
             
\maketitle

\begin{history}
\received{Day Month Year}
\revised{Day Month Year}
\end{history}

\begin{abstract}
In space-based gravitational wave detection, the estimation of far-field wavefront error of the distorted beam is the precondition for the noise reduction. 
Zernike polynomials is used to describe the wavefront error of the transmitted distorted beam.  
The propagation of a laser beam between two telescope apertures is calculated numerically. 
Far-field wavefront error is estimated with the absolute height of the peak-to-valley phase deviation between distorted Gaussian beam and a reference distortion-free Gaussian beam.  
The results show the pointing jitter is strongly related to the wavefront error. Furthermore, when jitter decreases 10 times from 100 to 10 nrad, wavefront error reduces for more than an order of magnitude.
In the analysis of multi-parameter minimization, the minimum of wavefront error tends to Z[5,3] Zernike in some parameter ranges.
Some Zernikes have a strong correlation with wavefront error of the received beam. 
When the aperture diameter increases at Z[5,3] Zernike, wavefront error is not monotonic and has oscillation.
Nevertheless, wavefront error almost remains constant with the arm length increasing from 10$^{-1}$ Mkm to 10$^3$ Mkm. When the arm length decreases for three orders of magnitude from 10$^{-1}$ Mkm to 10$^{-4}$ Mkm, wavefront error has only an order of magnitude increasing.
In the range of 10$^{-4}$ Mkm to 10$^3$ Mkm, the lowest limit of the wavefront error is from 0.5 fm to 0.015 fm, at Z[5,3] Zernike and 10 nrad jitter.
\end{abstract}
\ccode{PACS Nos.: 42.60.-v, 07.87.+v, 04.80.Nn}

\section{Introduction}\label{se:1}%============
There are 90 events of gravitational wave above 10Hz detected by the ground-based observatories from LIGO, VIRGO and KAGRA collaborations\cite{LIGOScientific:2021djp}. The gravitational wave observatory below 1 Hz is a space-based mission matching the longer baseline with the order of $10^6$ kms apart between the spacecrafts of gravitational wave detection\cite{Ruan:2020smc}. The less than 10 pm phase of the laser beams is shifted by the gravitational wave linked the spacecrafts, which is sensitive to the gravitational wave sources, such as super massive black hole binaries, the Galactic white dwarf binaries etc\cite{Jennrich:2009ti,Audley:2017drz}. 
For this level of precision, many measurement noises need to be suppressed to enhance the signal to noise ratio of gravitational wave sources. 
The basic one of the noise sources of the heterodyne interferometry in the spacecrafts, called tilt-to-length(TTL) coupling, is the coupling between an angular jitter of the interfering beams and the path length readout, which brings the extra optical path length into the interferometric measurements\cite{schuster2015vanishing}. 
TTL coupling not only affects the phase directly in geometric and non-geometric form, but also couples with wavefront errors\cite{robertson1997optics}. 
Although the transmitted beam obtains initial wavefront errors before propagation in space, the truncated wavefront has the less errors in the receiver aperture after the propagation over long distances\cite{Sasso:2018lja}.
Thus, the tight requirement on the phase stability of the received wavefront is the precondition of the noise reduction\cite{Thorpe:2009wg}. A detailed estimation of the received wavefront errors should be performed first. Moreover, for one of the major noises in ground-based observatories is also relevant to TTL coupling\cite{2005OExpr..13.7118M}. It is obvious that the lower limit of the received wavefront errors constrains the sensitivity to the detection of the weaker strain gravitational wave sources.

The TTL couplings between the wavefront misalignment and some low-order aberrations of the interfering beams are investigated analytically\cite{Sasso:2018lja}, which is also extended to higher order modes\cite{zhao2020tilt}. Utilizing a Zernike polynomial decomposition of wavefront error, the fast and transparent modeling techniques, that is derived from the neglecting a significant number of unimportant terms in the expansions of Zernike polynomials, for the purpose of estimating TTL noise associated with various wavefront error are introduced in the paper\cite{Weaver:2022btj}, which indicates that the certain combinations of these Zernikes were capable of producing noise far below the Laser Interferometer Space Antenna (LISA)\cite{Audley:2017drz} requirements\cite{Weaver:2022btj}. It is also found that the numerical simulation is used to describe the effect of an aberrated transmitting telescope on the light collected by the receiving telescope with Zernike modes\cite{Kenny:2020duu}. Via calculations of the wavefront aberrations in the far field, an end-to-end investigation of the measurement noise due to the interaction between the telescope jitters and wavefront aberrations are found in the paper\cite{Sasso:2019bit}.

In many papers,the wavefront error is estimated at the given parameters: 2.5 $ Mkm$ arm length and 300 mm aperture diameter etc, which are from LISA configuration.  
In this paper, we estimate the changes of far-field propagation wavefront error with the arm length, the aperture diameter and the pointing jitter, which are used generally for the gravitational wave detection.
We use Nijboer-Zernike Theory\cite{nijboer1947diffraction} to describe the diffraction propagation of the transmitted distorted beam. 
For the first time, the numerical calculation of the diffraction propagation between two telescope apertures is performed with the Gaussian Beam Decomposition(GBD)\cite{greynolds1986propagation} based on the numerical calculations. The numerical calculations are performed using the software package IfoCAD\cite{2012OptCo.285.4831W}.

The results of this paper show that the consistent wavefront error appears in the different methods between the Hermite–Gaussian modal decomposition\cite{Weaver:2022btj} and GBD at the same parameters. That justifies the GBD using for the diffraction propagation of laser beam. More informative results indicate that the wavefront error has no significant difference above 10$^4$km, but significant increase below 10$^4$km and is affected remarkably by the aperture diameter and the pointing jitter. With the aperture diameter varying, wavefront error of the different aperture diameters is not monotonic and has the oscillation. 

In Section \ref{se:2}, we introduce the TTL-Wavefront distortion coupling and the expression of distortion wavefront. 
In Section \ref{se:3}, there is detailed description about GBD used for the propagation of the laser beam as the sum of a series of fundamental Gaussian beams.
In Section \ref{se:4}, the numerical results are shown. In Section \ref{se:5}, the estimation of the far-field wavefront error is summarized as the conclusion.

\section{TTL-Wavefront distortion coupling and the expression of distortion wavefront} \label{se:2}%============
\begin{figure*}[htbp]
	\begin{center}
		\includegraphics[width=0.9\textwidth]{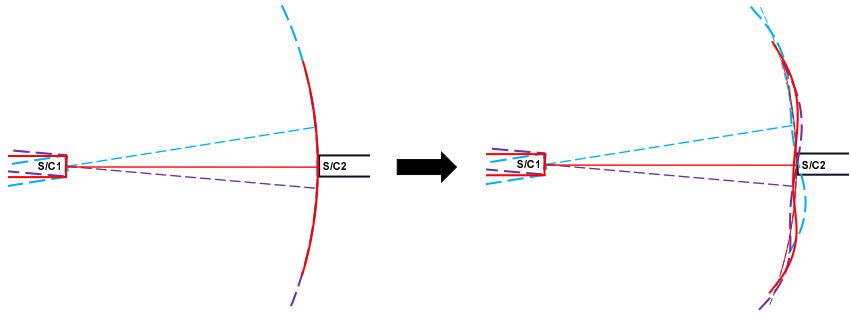}
	\end{center}
	\caption{Schematic diagram of how TTL-Wavefront distortion coupling occurs. (left) The ideal transmitted beam with the Jitter propagating into the telescope aperture of another spacecraft. (right) The propagation of the beam carrying wavefront errors. The truncated wavefront is transformed from the spherical wave(left) to the distorted one(right).}
	\label{TiltSetup}
\end{figure*}
In space-based gravitational wave detection, the ideal transmitted beam propagating into the telescope aperture of another spacecraft(S/C) is described as a truncated Gaussian beam\cite{10.1002/0471213748.ch3}. 
After the extremely far field propagation, the wavefront of the truncated Gaussian beam is approximated as a spherical wave. 
Thus, the received Top-hat beam has no an additional phase shift caused by the jitter of initial beam, which is shown in the left part of Fig. \ref{TiltSetup}.
However, through the local telescope system, the wavefront of transmitted beam is distorted usually by the various mechanisms.
Before being transmitted to another S/C, wavefront carries errors and is not an ideal beam. 
After far distance propagation, the shape of wavefront, which deviates from a spherical wave, is derived from the initial wavefront errors. That is shown in the right part of Fig. \ref{TiltSetup}. Generally, this mechanism is called as TTL-Wavefront distortion coupling, i.e., the jitter of transmitter brings the inconstant wavefront with the received one at the receiver. The measurement of a laser beam has the additional phase offset.

Frits Zernike and Bernard Nijboer developed a diffraction theory, which provides a description of the aberrated complex field in the image plane \cite{nijboer1947diffraction}. Through the aperture of telescope, the electric field is expressed as:

\begin{equation}
	E_a(r; 0)=E_0(r; 0)e^{i\varOmega_a},
		\label{wfegeneraldefinition}
\end{equation}
where $E_0(r; 0)$ is the amplitude of an assumed truncated circular Gaussian beam:
\begin{equation}
	E_0(r; 0)=
	\left\{
	\begin{aligned}
		e^{-r^2/w_0^2},&&\mbox{if $r\leq{r_a}$ } \\
		0,&&\mbox{if $r>{r_a}$ } 
	\end{aligned}
	\right.
\end{equation}
$w_0$ is the waist radius of Gaussian beam and $r_a$ is the aperture radius of telescope. $\varOmega_a$ in \eqref{wfegeneraldefinition} is the total phase departure, which is formed by a set of Zernike polynomials:

	\begin{equation}
	\begin{aligned}
		\varOmega_a(\rho,\theta)=\sum_{n=0}^{N}\sum_{m=-n}^{n} c^m_n\,Z^m_n(\rho,\theta),\\
	\mbox{where }Z^m_n(\rho,\theta)=
	\left\{
	\begin{aligned}
		R^m_n(\rho)\cos\left(m\theta\right),&&\mbox{if $m\geq0$ } \\
		R^{-m}_n(\rho)\sin\left(-m\theta\right),&&\mbox{if $m<0$ }
	\end{aligned}
	,\right. \\
	R_n^{|m|}(\rho)=(-1)^{(n-|m|)/2}\rho^{|m|}P_{(n-|m|)/2}^{(|m|,0)}(1-2\rho^2),
	\end{aligned}
	\label{AberrationDefinition}
	\end{equation}
where $c^m_n$ represents coefficients and $Z^m_n$ are Zernike Polynomals written with Noll indexing\cite{Noll:76}. $\rho=r/r_a$ is restricted to unit disk(0$\leq\rho\leq$1), and $\theta$ is the azimuth. $n-m\geq0$ and is even. Zernike Polynomals $Z^m_n(\rho,\theta)$ is a circle polynomials.
$R^m_n(\rho)$ is the radial polynomials.
where $P_{k}^{(\alpha,\beta)}$ is the Jacobi polynomial of degree k, and $R^m_n(\rho)$ satisfies the orthogonality relation:
\begin{equation}
	\int_0^1R_n^{|m|}(\rho)R_{n`}^{|m|}(\rho)\rho{d}\rho=\frac{\delta_{n,n`}R_n^{|m|}(1)}{2(n+1)},
\end{equation}

When the propagation distance is very far for space-based gravitational wave detection, and the pointing jitter of S/C is limited to $10^{-8}rad/\sqrt{Hz}$ level, the diffraction integral satisfies the conditions for Fraunhofer diffraction:
\begin{equation}
	E(r, \psi, z)=\frac{e^{ikz}e^{\frac{ik}{2z}r^2}}{i\lambda{z}}{r_a}^2
	\int_0^1\int_0^{2\pi}e^{-\rho^2}e^{i\varOmega_a(\rho,\theta)}e^{-iv\rho\cos{(\theta-\psi)}}\rho{d}\rho{d}\theta,
\label{FraunhoferAberration1}
\end{equation}
where $w_0=r_a$ is assumed and $v=\frac{k}{{z}}{r_a}r$. 

The expression\eqref{FraunhoferAberration1} is computed numerically, which is performed by C++ code in the IfoCAD package\cite{2012OptCo.285.4831W}.

In the expression\eqref{FraunhoferAberration1}, $E(r, \psi, z)$ is expressed as $A(x, y, z)e^{i\Theta(x, y, z)}$, where A(x, y, z) is the amplitude of $E(r, \psi, z)$ and  
$\Theta(x, y, z)$ is the phase of Wavefront:
\begin{equation}
\begin{aligned}
&E(r, \psi, z)=A(x, y, z)e^{i\Theta(x, y, z)}, \\
&\mbox{where } 
\left\{
	\begin{aligned}
x=rcos(\psi),\\ y=rsin(\psi).
\end{aligned}\right.
\end{aligned}
\end{equation}
The phase difference $\delta\Theta(x, y, z)$ of wavefront is between the distorted Gaussian beam $E(r, \psi, z)$ and the given reference distortion-free Gaussian beam $E(r, \psi, z)_0$.
Expressions are in the following:
\begin{equation}
\begin{aligned}
\delta\Theta(x, y, z)=\Theta(x, y, z)-\Theta(x, y, z)_0.
\end{aligned}
\label{deltaTheta}
\end{equation}

\section{Gaussian Beam Decomposition} \label{se:3}%============
As the mixture of zernike polynomials increases, the expression\eqref{FraunhoferAberration1} becomes more complex. 
Via neglecting a significant number of unimportant terms in the expansions of Zernike polynomials and Hermite–Gaussian modal decomposition based propagation of an approximation in the initial beam, TTL noise associated with various wavefront error is estimated in the paper\cite{Weaver:2022btj}. Furthermore, the beam decomposition methods are required for more conveniently simulating the diffraction of an aberrated or distorted Gaussian beam. 
\begin{figure*}[htbp]
	\begin{center}
		\includegraphics[width=1\textwidth]{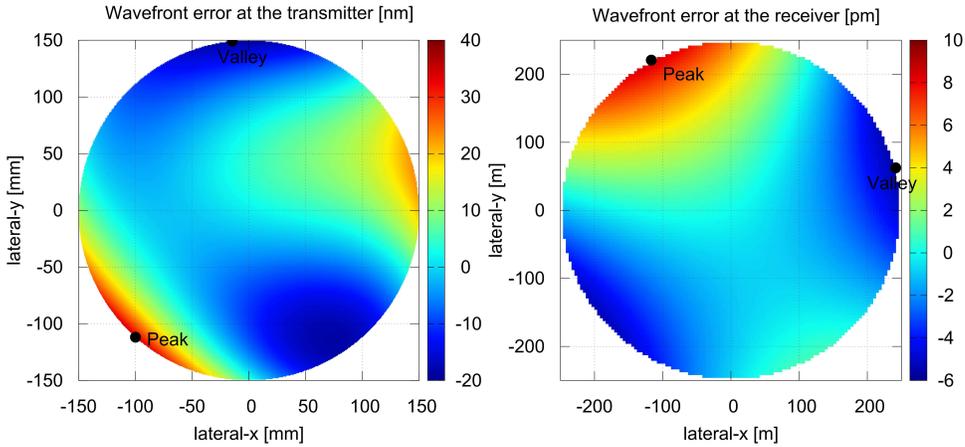}
	\end{center}
	\caption{Wavefront at the transmitter(left) and the wavefront error(right) at the receiver. The coefficients of Zernike polynomials for the numerical calculation are listed in Table \ref{SassoAberrationParameter}. The transmitted wavefront has $\lambda/20$ peak-to-valley deviation from a plane. The numerical calculated values of Peak and Valley are 32.9119 nm and -20.2904nm respectively. The wavefront error described by PV in the right graph is about 14.1pm. The numerical calculated values of Peak and Valley are 8.49655 pm and -5.61987 pm respectively.}
	\label{SassoExample}
\end{figure*}

In this paper, the Gaussian Beam Decomposition(GBD) is used for the beam profile decomposition to compute the diffraction propagation including propagation between the optical components. The basic idea of GBD is to describe the propagation of a beam as the sum of a series of fundamental Gaussian beam propagation. These Gaussian beams are distributed on a grid, also called grid beams. Each grid beam propagates following the ABCD law and has its own intersection point with optical components. Therefore, GBD can also be considered as one kind of "fat rays" tracing. GBD is firstly proposed by Greynolds in 1986\cite{greynolds1986propagation}, and has been further developed in recent years. 
The further details of GBD and its computation process can be found in \ref{se:7}. 
This paper focuses on the diffraction propagation of aberrated Gaussian beam by using GBD. We program C++ code to perform these calculations numerically. The code is included into IfoCAD\cite{2012OptCo.285.4831W} package for the first time.

\begin{table}
\tbl{The List of the parameters for distorted circular Gaussian beam clipped by a circular aperture centered in the beam waist.}{
\centering
		\begin{tabular}{|l|l|l|}
			\hline
			\textbf{Parameter} & \textbf{Description} &  \textbf{Value}  \\ 
			\hline
			$\lambda$ & wavelength & 1064 nm \\ 
			\hline
			$\omega_0$ & beam waist & 150 mm \\ 
			\hline
			$z_0$ & distance to the waist & 0  \\ 
			\hline
			$P$ & power & 1W  \\ 
			\hline
			$ds$ & propagation distance & 2.5 Mkm  \\ 
			\hline
			$r_a$ & aperture radius &  150 mm \\ 
			\hline
		\end{tabular}
		\label{SassoBeamParameter}}
\end{table}

By means of GBD we can obtain the radial electric field distribution of the transmitted beam at the far field after propagation. 
We firstly perform the numerical calculations based on the same parameters as C.P Sasso's work in the paper\cite{Sasso:2018lja}, to verify the validity of GBD method. These parameters are listed in Table \ref{SassoBeamParameter}. The wavefront error generated randomly is constrained to $\lambda/20$ peak-to-valley deviation from a plane. The radial plane radius of the received beam is determined by the jitter of S/C:
\begin{equation}
	r=L \cdot a,
\label{RadialPlane}
\end{equation}
when $L$ is  2.5 Mkm arm length, and $a$ is 100 nrad jitter, the receiving plane corresponds to a circular plane with 250 m radius. On this plane, we compute the phase difference $\delta\Theta$ between distorted Gaussian beam and a reference distortion-free Gaussian beam with the expression \ref{deltaTheta}. The peak-to-valley phase difference are calculated as the following:
\begin{equation}
	PV=\frac{{|\delta{\Theta}}_{Peak}|+|{\delta{\Theta}}_{Valley}|}{2\pi}{\lambda},
\label{RadialPlane}
\end{equation}
The best-fit coefficients of Zernike polynomials in expression \eqref{AberrationDefinition} for the transmitted beam are shown in Table \ref{SassoAberrationParameter}. The results of relevant calculation are shown in Fig.\ref{SassoExample}, where the left is the wavefront error at the transmitter, and the right one is wavefront error at the receiving plane. It is shown that PV=14.116 pm is consistent with the result in the paper\cite{Sasso:2018lja}.

\begin{table}
\tbl{The best-fit Zernike Coefficients of distortion, which come from fitting Zernike polynomials to the wavefront at receiver.}{
\centering
		\begin{tabular}{c|c c c c c c}
			\hline
			$Z^m_n$ & $Z^0_0$ &  $Z^{-1}_1$ &  $Z^{1}_1$ &  $Z^{-2}_2$ &  $Z^{0}_2$ &  $Z^{2}_2$  \\ 
			\hline
			$ c^m_n$ & 0. & 0. & 0. & 0.0967293  &  0.00795719 & 0.0662892  \\
			\hline
			$  $ & $Z^{-3}_3$ &  $Z^{-1}_3$ &  $Z^{1}_3$ &  $Z^{3}_3$ &  $Z^{0}_4$ &     \\ 
			\hline
			$  $ & 0.0130634 & -0.0565416 & -0.0225312 & 0.0491065  & 0.0164262   &      \\
			\hline
		\end{tabular}
\label{SassoAberrationParameter} }
\end{table}

\section{Numerical results} \label{se:4}%============
Based on the given parameters including propagation distances in the context, the numerical calculation of wavefront(WF) error have been verified for consistency. In order to analyze deeply the lower limit of far-field wavefront error, the different parameters: the arm length, the aperture diameter and the pointing jitter are also considered.  
\begin{figure*}[htbp]
	\begin{center}
		\includegraphics[width=0.9\textwidth]{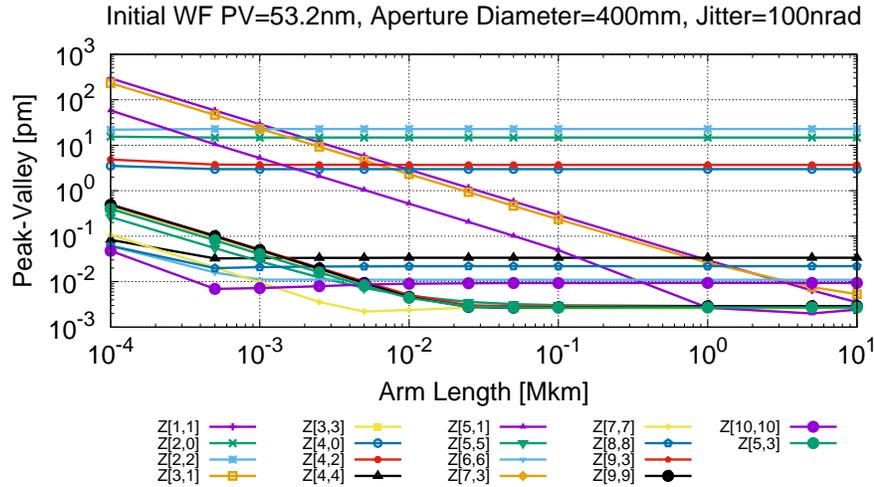}
	\end{center}
	\caption{At 100 nrad jitter, PV values of the wavefront error of distorted gaussian beam at the receiver after propagation, which are relevant to the different Zernikes. The initial wavefront error is constrained to 53.2 nm. The diameter of telescope aperture is 400 mm.}
	\label{PV-100nrad} 
\end{figure*}

\begin{figure*}[htbp]
	\begin{center}
		\includegraphics[width=0.9\textwidth]{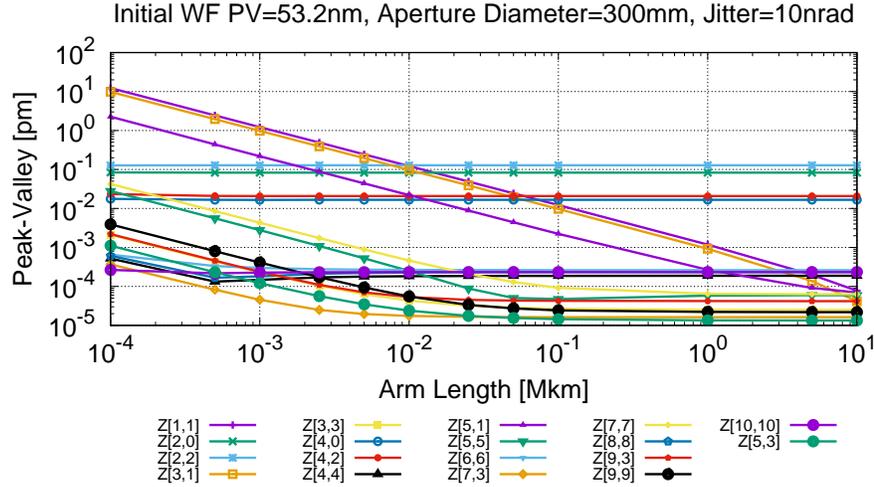}
	\end{center}
	\caption{Similar results to Fig. \ref{PV-100nrad}. differently this figure is relevant to 10 nrad jitter and 300 mm aperture diameter.}
	\label{PV-10nrad} 
\end{figure*}

Firstly, the pointing jitter is chosen as 100 nrad, which is used at LISA configuration. 
Besides, the invariant parameters are initial wavefront error $\lambda/20$  and aperture diameter 400 mm.
The propagation distance expressed as arm length is in the range of $10^{-4}$ to $10^{3}$ Mkm for space-based gravitational wave detection. 
The Zernikes $Z^m_n$ are scanned by changing m and n values. 
Based on the parameters, the far-field wavefront error is calculated by IfoCAD\cite{2012OptCo.285.4831W} package including GDB method in the context. 
The numerical results are shown in Fig. \ref{PV-100nrad}. 

As is seen in Fig. \ref{PV-100nrad}, the PV values of wavefront error related to the different Zernikes have the remarkable difference for the several orders of magnitude picometer. When Zernikes $Z^m_n$ are scanned by changing m and n values, it is found that the main source of wavefront errors is relevant to low-order aberration (n$<5$) and higher-order aberrations become more obvious in shorter arm lengths than long arm. The lower limit of wavefront error is in the range of 0.1 to 0.002 pm related to the arm length from 10$^{-4}$ to 10 Mkm.

Secondly, the pointing jitter is changed as 10 nrad decreased an order than the above value. The aperture diameter is chosen as 300mm. 
As is seen in  Fig. \ref{PV-10nrad} that with a smaller jitter, the PV values of wavefront error are depressed by 1 to 2 orders of magnitude, covering the whole arm length range. Furthermore, the trend of PV values has not much changes with arm length increasing. 
In the arm length range of 0.1 to 10 Mkm, the main sources of wavefront errors are relevant to Z[2,0], Z[2,$\pm$2], Z[4,0] and Z[4,$\pm$2], corresponding to Defocus, Astigmatism, Primary Spehrical, and 2nd Astigmatism, respectively. In this case, the lower limit of wavefront error is from 0.5 to 0.015 fm.

\begin{figure*}[htbp]
	\begin{center}
		\includegraphics[width=0.9\textwidth]{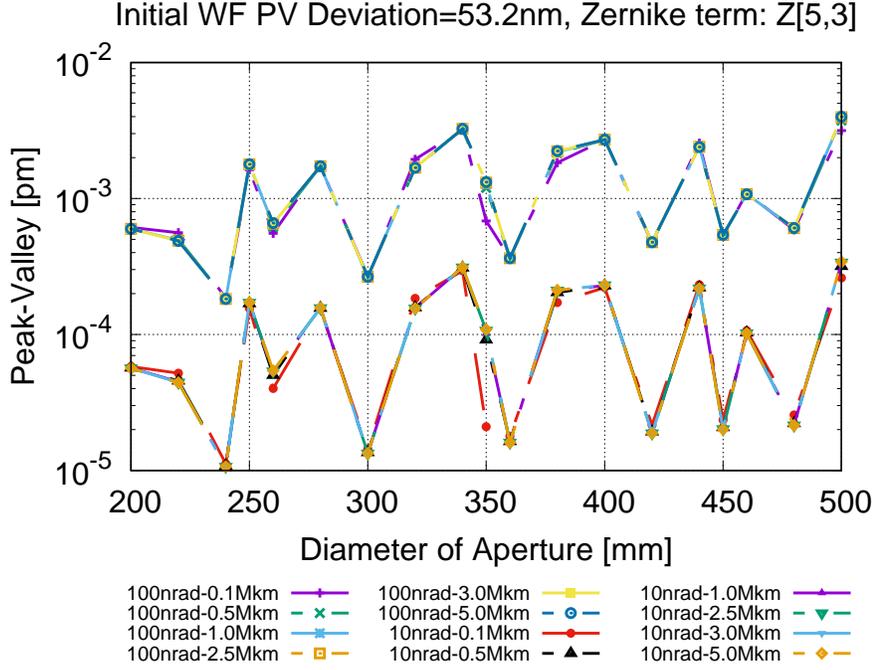}
	\end{center}
	\caption{At Zernike Z[5,3], PV values of the Wavefront error of propagated beam with the telescope aperture diameter increasing at the transmitter. }
	\label{Z53D} 
\end{figure*}
Finally, the effect of different telescope apertures is considered. The aperture diameters vary from 200 to 500 mm. The 100 nrad and 10 nrad jitter are both used.
Referring to the results from Fig. \ref{PV-100nrad} and \ref{PV-10nrad}, the chosen Z[5,3] is almost relevant to the least PV values of wavefront error among the Zernikes in the arm length range of 0.1 to 10 Mkm. 
As is seen in Fig. \ref{Z53D}, with the aperture diameter increasing, the wavefront error is not monotonic and has the oscillation. Two groups of the broken lines are relevant to the different jitters. The lower lines are relevant to the 10 nrad jitter. The lowest limit of wavefront error is near 0.01 fm consistent with the results in Fig. \ref{PV-10nrad}. As an inference, the aperture diameter selection is not easy to be fixed at the range for reducing wavefront error.

\begin{figure*}[htbp]
	\begin{center}
		\includegraphics[width=0.48\textwidth]{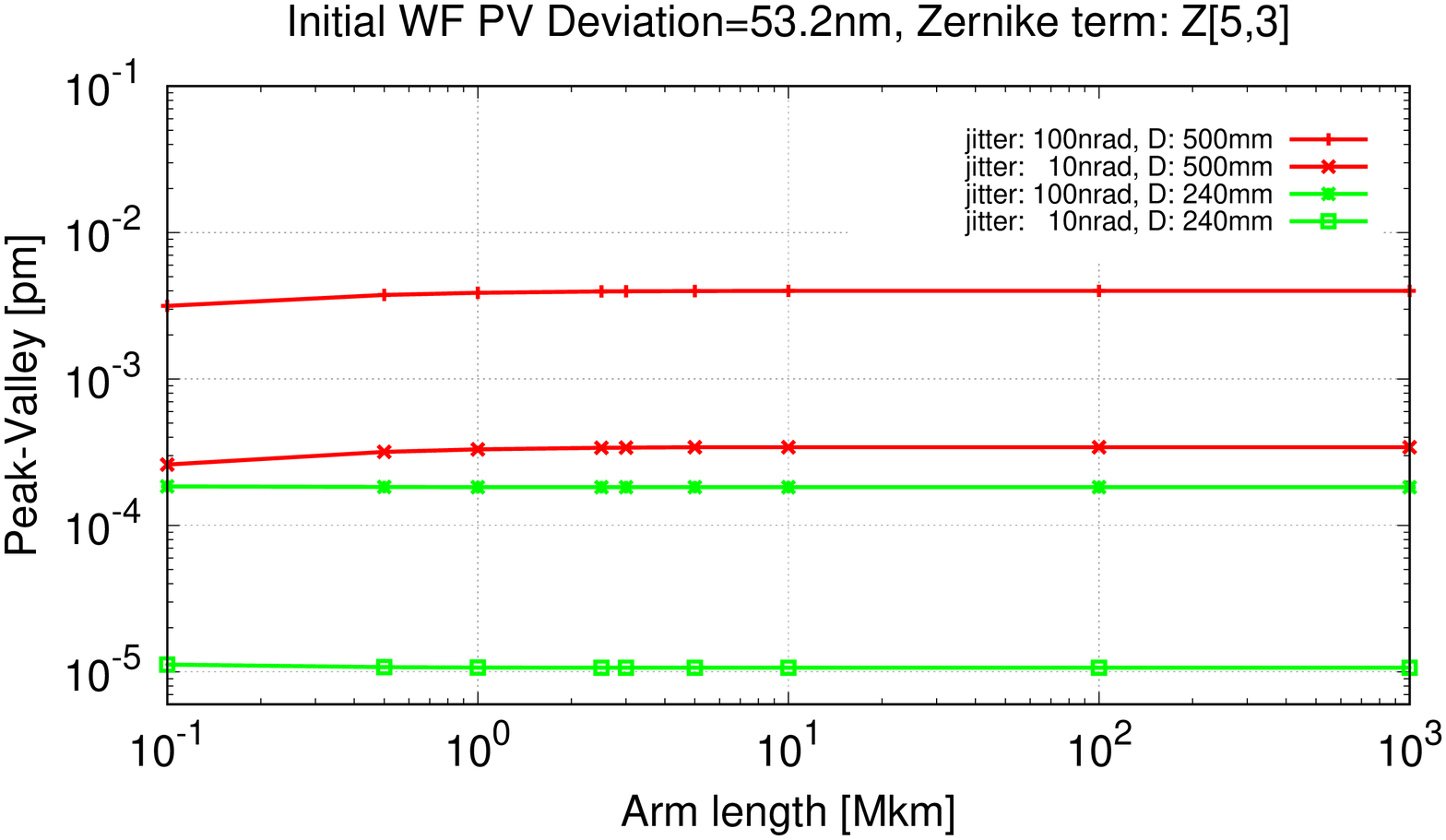}\includegraphics[width=0.48\textwidth]{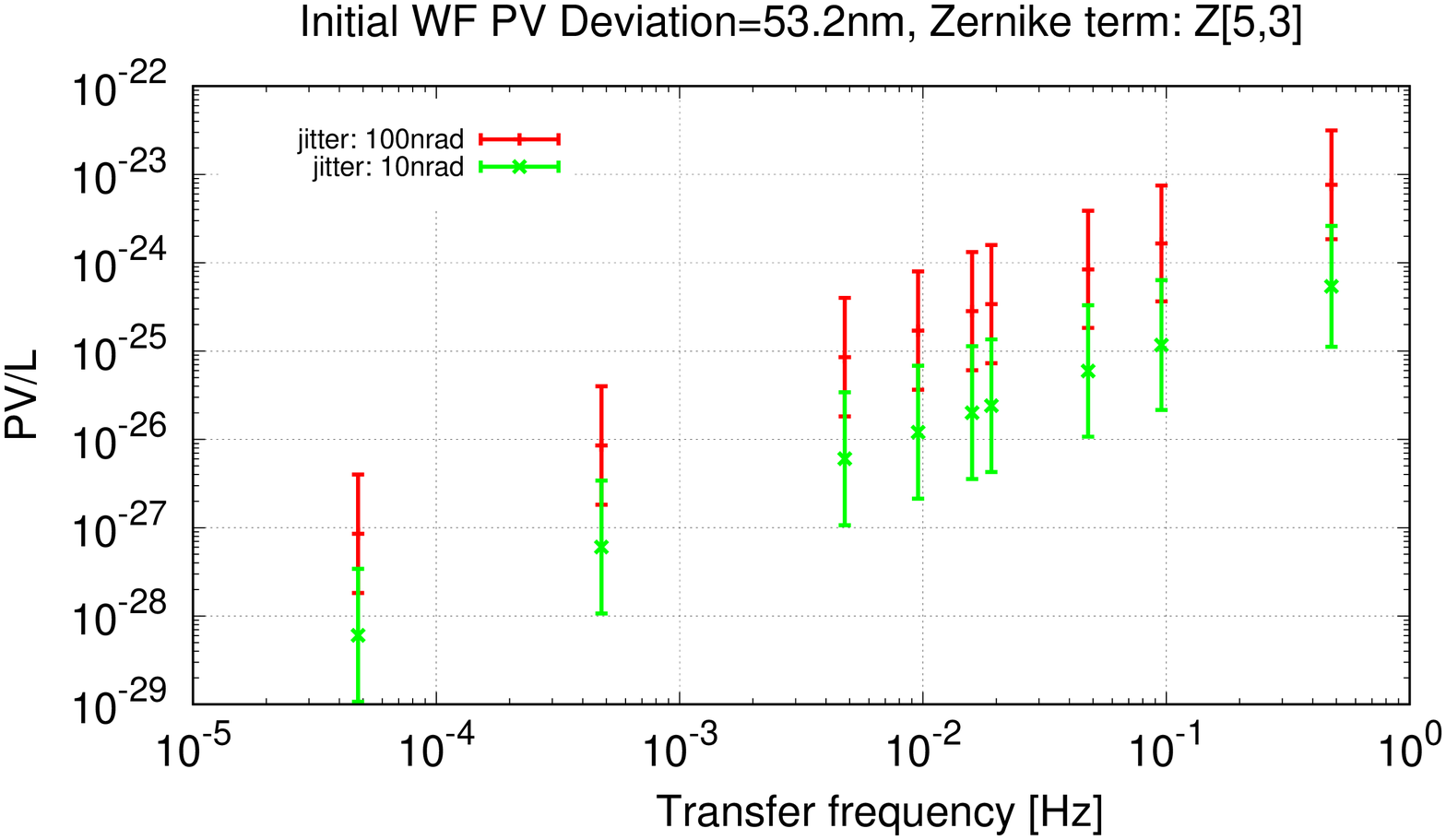}
	\end{center}
	\caption{(left) PV values relevant to the arm length from 10$^{-1}$ to 10$^3$ Mkm. (right) the calculated PV/L values with the transfer frequency increasing. }
	\label{LPV} 
\end{figure*}

When the distance of laser beam propagation is very long, the change of wavefront error is tiny. That result is shown in the left of Fig. \ref{LPV}. Two group of the lines are relevant to the 100 nrad and 10 nrad jitter respectively. wavefront error is very slightly increasing in the arm length range of 10$^{-1}$ Mkm to 10$^3$ Mkm. The different aperture diameters of 240 and 500 mm bring wavefront error an order of magnitude increasing.

The magnitude of the gravitational wave is the strain h=$\delta L/L$ characterized the change of the proper distance between the space-time points.
In the heterodyne interferometry, the wavefront plane of laser beam is like the scale of a ruler to measure the distance. 
The wavefront error means the scale error of a ruler. Thus, PV/L values have effect on the measured magnitude of gravitational wave. 
The relation between the best arm length $L$ of S/Cs and the detectable frequency $f_{*}$ of gravitational wave source is expressed as the transfer frequency: $f_{*}=\frac{C}{2{\pi}L}$(C is the light speed)\cite{Prince:2002hp}. 
Thus, the change of PV/L with arm length increasing is transformed into the transfer frequency decreasing. 
These calculations are shown in the right of Fig. \ref{LPV}. The upper and lower bound of PV/L values are relevant to the aperture diameters of 500 mm and 240 mm. 
As is seen in the right of Fig. \ref{LPV}, the V/L values increase with transfer frequency increasing and lowest limit of PV/L is relevant to the 10 nrad jitter. 
Thus, as an inference, the decrease of the wavefront error brings the sensitivity to gravitational wave sources increasing.

\section{Conclusion} \label{se:5}%============
In space-based gravitational wave detection, TTL-Wavefront distortion coupling is one of major noise sources. 
In the paper, we estimate the wavefront error, which is derived from the transmitted beam is distorted usually by the various mechanisms. 
The initial wavefront error generated randomly is constrained to $\lambda/20$.
Zernike polynomials is used for the description of the aberrated wavefront of the transmitted beam.
In the analysis of multi-parameter minimization, the minimum of wavefront error tends to Z[5,3] Zernike in some parameter ranges. 
Some Zernikes have a strong correlation with wavefront error of the received beam.
By means of Gaussian Beam Decomposition, we obtain the radial electric field distribution of the transmitted beam at the far field after propagation. 
The phase difference between distorted Gaussian beam and a reference distortion-free Gaussian beam  is calculated by the peak-to-vallet values.

In the paper, the variant parameters: the arm length, the aperture diameter and the pointing jitter are considered. 
The pointing jitter is very sensitive to the wavefront error, i.e., 10 times decrease from 100 to 10 nrad brings the far-field wavefront error more than an order of magnitude reduction.
With the aperture diameter varying, wavefront error of the different aperture diameters are not monotonic and have the oscillation.  
It is implied that the aperture diameter is not easy to be fixed at the range for reducing far-field wavefront error.
Wavefront error is very slightly increasing from 10$^{-1}$ Mkm to 10$^3$ Mkm arm length at Z[5,3] Zernike and 10 nrad jitter. 
In the range of 10$^{-4}$ Mkm to 10$^3$ Mkm, the lowest limit of wavefront error changes from 0.5 fm to 0.015 fm, at Z[5,3] Zernike and 10 nrad jitter.
All the results imply that the decrease of the wavefront error brings the increase of the sensitivity to gravitational wave sources.

\section{Acknowledgements}%============
We thank Gudrun Wanner from Max Planck Institute for Gravitational Physics (Albert Einstein Institute) for the great contribution to the improvement of the quality of this paper. 
We thank Yun-Kau Lau for the useful discussions. 
This work has been supported in part by the National Key Research and Development Program of China under Grant No.2020YFC2201501, the National Science Foundation of China (NSFC) under Grants No. 12147103 (special fund to the center for quanta-to-cosmos theoretical physics), No. 11821505, the Strategic Priority Research Program of the Chinese Academy of Sciences under Grant No. XDB23030100, and the Chinese Academy of Sciences (CAS). In this paper, the part of the numerical computation is finished by TAIJI Cluster.

%\noindent
%Corresponding authors: $^{*}$ hbjin@bao.ac.cn

\appendix%============
\section{Gaussian Beam Decomposition(GBD)} \label{se:7}
\begin{figure*}[htbp]
	\begin{center}
		\includegraphics[width=0.7\textwidth]{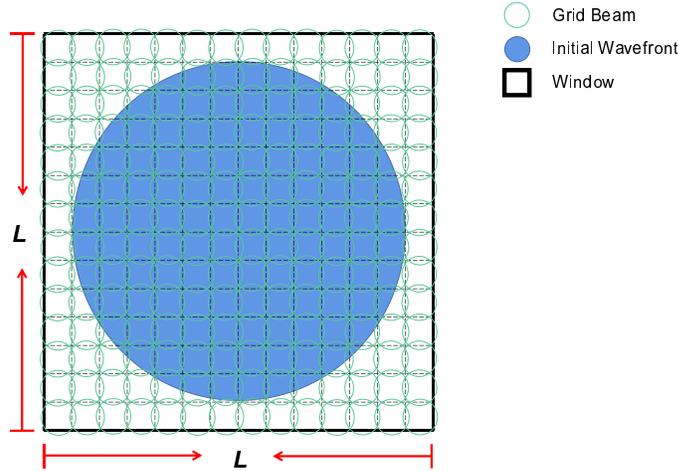}
	\end{center}
	\caption{A simple illustration of GBD method. The green one is fundamental grid Gaussian beam, which is arranged in window. The window in the graph is chosen to be square. The blue one is the wavefront to be decomposed. }
	\label{GBDCIllustration}
\end{figure*}

The basic idea of GBD is to describe the propagation of a beam as the sum of a series of fundamental Gaussian beams. The propagation of Gaussian beam is easy to be computed. Therefore, the implementation of GBD is to find the coefficient of each chosen fundamental Gaussian beams, and superimpose these fundamental beams to reconstruct the electric field. A illustration is shown in Fig. \ref{GBDCIllustration}.
 By taking enough sampling points in space, the problem can be expressed as solving a system of linear equations by matrix:
\begin{equation}
	B\vec{c} = \vec{W_f}
\end{equation}
where $W_f$ represents the complex electric field sampled from the wavefront to be decomposed. $\vec{c}$ is the coefficient vector of fundamental Gaussian beams, and B is the set of ${N}{\times}{N}$ fundamental Gaussian beams in space: 	

\begin{equation}
		\begin{array}{ll}
			\vec{c} = 
			\begin{pmatrix}
				c_{1}\\
				c_{2}\\
				\vdots \\

				\vdots \\
				c_{N*N} \\
			\end{pmatrix}, \
			\vec{W_{\emph{f}}} = 
			\begin{pmatrix}
				w_{1}\\
				w_{2}\\
				\vdots \\
				w_{N*N} \\
			\end{pmatrix} \,.			
		\end{array}
	\end{equation}

	\begin{equation}
		B =
		\begin{bmatrix}
			b_{1}(\textbf{\emph{x}}_1) &  b_{2}(\textbf{\emph{x}}_1) & \cdots  & b_{N*N}(\textbf{\emph{x}}_1)   \\
			b_{1}(\textbf{\emph{x}}_2) &  b_{2}(\textbf{\emph{x}}_2) & \cdots  & b_{N*N}(\textbf{\emph{x}}_2)   \\
			\vdots & \cdots  & \ddots   & \vdots  \\
			b_{1}(\textbf{\emph{x}}_{N*N}) & b_{2}(\textbf{\emph{x}}_{N*N}) & \cdots & b_{N*N}(\textbf{\emph{x}}_{N*N})  \\
		\end{bmatrix}
	\end{equation}
Fundamental Gaussian beams are chosen on a grid of ${N}{\times}{N}$ in the plane area, which is called window. 
Fundamental Gaussian beams are placed on the center of each grid, also known as grid beams.
On the window, there are various options for grid, the simplest being an square. The size of the window should be large enough to cover the entire beam spot, otherwise the decomposition accuracy will be highly affected because of information loss. It should be noted that choosing grid in the plane is not essential, but is a simple and practical way. In the paper\cite{worku2018decomposition}  methods of taking grid points was discussed on curved surfaces. 

Considering the uniqueness of the solution to the system of matrix linear equations, matrix B should be determined. Therefore the number of space data points $\textbf{\emph{x}}_{i}$ should be no less than the number of grid beams ${N}{\times}{N}$. Another relevant parameter for the chosen of grid beams is the waist factor $f_{{\rm ws}}$. It determines both the waist of all grid beams and the tightness of the arrangement between grid beams:

	\begin{equation}
		w_{0g} = \frac{D}{f_{{\rm ws}}},
		\label{fws}
	\end{equation} 
where D is the interval between two adjacent grid beams, determined by window size and the grid beam number:
	\begin{equation}
	D = \frac{L}{N}.
	\label{griddistance}
\end{equation} 
In this paper, the relevant parameters of GBD chosen in simulations are all based on Table \ref{GBDParameter}.

\begin{table}
\tbl{The List of the parameters for GBD.}{
\centering
		\begin{tabular}{|l|l|l|}
			\hline
			\textbf{Parameter} & \textbf{Description} &  \textbf{Value}  \\ 
			\hline
			$n$ & grid number of the GBD & 200$\times$200\\ 
			\hline
			$L$ & window size of the GBD & $\frac{8}{3} {r_a}$ \\
			\hline
			$f_{ws}$ & waist factor of the GBD&  0.75\\
			\hline
			$s$ & grid shape of the GBD &   square   \\
			\hline
		\end{tabular}
		\label{GBDParameter}}
\end{table}

\bibliography{reference,library}

\begin{thebibliography}{10}

\bibitem{LIGOScientific:2021djp}
{The LIGO Scientific Collaboration}, {the Virgo Collaboration} and {the KAGRA
  Collaboration} (Nov 2021) \href{http://arxiv.org/abs/2111.03606}{{\ttfamily
  arXiv:2111.03606}}.

\bibitem{Ruan:2020smc}
W.-H. Ruan, C.~Liu, Z.-K. Guo, Y.-L. Wu and R.-G. Cai, {\em Nature Astronomy}
  {\bf 4} (Feb 2020) 108, \href{http://arxiv.org/abs/2002.03603}{{\ttfamily
  arXiv:2002.03603}}.

\bibitem{Jennrich:2009ti}
O.~Jennrich, {\em Classical and Quantum Gravity} {\bf 26} (Aug 2009)   153001.

\bibitem{Audley:2017drz}
P.~Amaro-Seoane, H.~Audley, S.~Babak, J.~Baker, E.~Barausse, P.~Bender,
  E.~Berti, P.~Binetruy, M.~Born, D.~Bortoluzzi, J.~Camp, C.~Caprini,
  V.~Cardoso, M.~Colpi, J.~Conklin, N.~Cornish, C.~Cutler, K.~Danzmann,
  R.~Dolesi, L.~Ferraioli, V.~Ferroni, E.~Fitzsimons, J.~Gair, L.~G. Bote,
  D.~Giardini, F.~Gibert, C.~Grimani, H.~Halloin, G.~Heinzel, T.~Hertog,
  M.~Hewitson, K.~Holley-Bockelmann, D.~Hollington, M.~Hueller, H.~Inchauspe,
  P.~Jetzer, N.~Karnesis, C.~Killow, A.~Klein, B.~Klipstein, N.~Korsakova,
  S.~L. Larson, J.~Livas, I.~Lloro, N.~Man, D.~Mance, J.~Martino, I.~Mateos,
  K.~McKenzie, S.~T. McWilliams, C.~Miller, G.~Mueller, G.~Nardini,
  G.~Nelemans, M.~Nofrarias, A.~Petiteau, P.~Pivato, E.~Plagnol, E.~Porter,
  J.~Reiche, D.~Robertson, N.~Robertson, E.~Rossi, G.~Russano, B.~Schutz,
  A.~Sesana, D.~Shoemaker, J.~Slutsky, C.~F. Sopuerta, T.~Sumner, N.~Tamanini,
  I.~Thorpe, M.~Troebs, M.~Vallisneri, A.~Vecchio, D.~Vetrugno, S.~Vitale,
  M.~Volonteri, G.~Wanner, H.~Ward, P.~Wass, W.~Weber, J.~Ziemer and P.~Zweifel
  (Feb 2017) \href{http://arxiv.org/abs/1702.00786}{{\ttfamily
  arXiv:1702.00786}}.

\bibitem{schuster2015vanishing}
S.~Schuster, G.~Wanner, M.~Tr{\"o}bs and G.~Heinzel, {\em Applied optics} {\bf
  54}  (2015) 1010.

\bibitem{robertson1997optics}
D.~Robertson, P.~McNamara, H.~Ward and J.~Hough, {\em Classical and Quantum
  Gravity} {\bf 14}  (1997)   1575.

\bibitem{Sasso:2018lja}
C.~P. Sasso, G.~Mana and S.~Mottini, {\em Classical and Quantum Gravity} {\bf
  35} (Sep 2018)   185013.

\bibitem{Thorpe:2009wg}
J.~I. Thorpe, {\em Classical and Quantum Gravity} {\bf 27} (Apr 2010)   084008,
  \href{http://arxiv.org/abs/0911.3175}{{\ttfamily arXiv:0911.3175 [gr-qc]}}.

\bibitem{2005OExpr..13.7118M}
G.~Mueller, {\em Optics Express} {\bf 13} (Sep 2005)   7118.

\bibitem{zhao2020tilt}
Y.~Zhao, J.~Shen, C.~Fang, H.~Liu, Z.~Wang and Z.~Luo, {\em Optics Express}
  {\bf 28}  (2020) 25545.

\bibitem{Weaver:2022btj}
A.~J. Weaver, G.~Mueller and P.~J. Fulda, {\em Classical and Quantum Gravity}
  {\bf 39} (Oct 2022)   195016.

\bibitem{Kenny:2020duu}
F.~Kenny and N.~Devaney, {\em Classical and Quantum Gravity} {\bf 38} (Feb
  2021)   035010.

\bibitem{Sasso:2019bit}
C.~P. Sasso, G.~Mana and S.~Mottini, {\em Optics Express} {\bf 27} (Jun 2019)
  16855.

\bibitem{nijboer1947diffraction}
B.~R.~A. Nijboer, {\em Physica} {\bf 13}  (1947) 605.

\bibitem{greynolds1986propagation}
A.~W. Greynolds, { Propagation of generally astigmatic gaussian beams along
  skew ray paths}, in {\em Diffraction Phenomena in Optical Engineering
  Applications\/},  (1986), pp. 33--51.

\bibitem{2012OptCo.285.4831W}
G.~Wanner, G.~Heinzel, E.~Kochkina, C.~Mahrdt, B.~S. Sheard, S.~Schuster and
  K.~Danzmann, {\em Optics Communications} {\bf 285} (Nov 2012) 4831.

\bibitem{10.1002/0471213748.ch3}
{Beam Optics}, in {\em Fundamentals of Photonics\/},  (John Wiley {\&} Sons,
  Inc., New York, USA, 1991), New York, USA, ch.~3, pp. 80--107.

\bibitem{Noll:76}
R.~J. Noll, {\em Journal of the Optical Society of America} {\bf 66} (Mar 1976)
    207.

\bibitem{Prince:2002hp}
T.~A. Prince, M.~Tinto, S.~L. Larson and J.~W. Armstrong, {\em Physical Review
  D} {\bf 66} (Dec 2002)   122002,
  \href{http://arxiv.org/abs/0209039}{{\ttfamily arXiv:0209039 [gr-qc]}}.

\bibitem{worku2018decomposition}
N.~G. Worku, R.~Hambach and H.~Gross, {\em JOSA A} {\bf 35}  (2018) 1091.

\end{thebibliography}
\bibliographystyle{ws-ijmpd}
\makeatletter
\end{document}